\documentclass[useAMS,usenatbib]{mn2e}

\usepackage{graphicx}
\usepackage{amsmath}
\usepackage{amssymb} 
\title[The Outer Halo Globular Clusters of M31]{The Outer Halo Globular
Clusters of M31\thanks{Based on observations obtained at the W. M. Keck
Observatory, which is operated as a
scientific partnership among the
California Institute of Technology, the University of California, and the
National Aeronautics and Space Administration.}\\}
\author[A. Alves-Brito et al. 2009]{Alan Alves-Brito$^{1}$\thanks{E-mail:
abrito@astro.swin.edu.au (AAB)}, Duncan A. Forbes$^{1}$, Jon T. Mendel$^{1}$,
George K.T. Hau$^{1}$ \\\\
\normalfont{\LARGE{and Michael T. Murphy$^{1}$}}\\
$^{1}$Centre for Astrophysics and Supercomputing, Swinburne University of
Technology, Hawthorn, Victoria 3122, Australia\\
}
\begin{document}

\date{Accepted 1988 December 15. Received 1988 December 14; in original form 1988 October 11}

\pagerange{\pageref{firstpage}--\pageref{lastpage}} \pubyear{2002}

\maketitle

\label{firstpage}

\begin{abstract}

We present Keck/HIRES spectra of 3
globular clusters in the outer halo of M31, at
projected distances beyond $\approx$ 80 kpc from M31.
The measured recession velocities for all 3 globular clusters confirm their
association with                                                                            
the globular cluster system of M31. 
We find evidence for a declining velocity dispersion with radius for the
globular cluster system.
Their measured internal velocity
dispersions,             
derived virial masses and mass-to-light ratios are consistent               
with those for the bulk of the M31 globular cluster system. We derive old                     
ages and metallicities which indicate that all 3 belong to the                  
metal-poor halo globular cluster subpopulation. We find indications that the                  
radial gradient of the mean metallicity of the globular cluster system interior
to 50 kpc flattens in the outer regions, however                       
it is still more metal-poor than the corresponding field stars at               
the same (projected) radius.                     

\end{abstract}
 

\section{Introduction}

One of the great unanswered questions in astrophysics is the formation and
evolution of galaxies in the Universe. Since M31 is our closest spiral galaxy,
located at a distance of $\sim$ 780 kpc (Holland 1998), it is considered an
ideal astrophysical laboratory to test the different ideas
about galaxy formation and evolution.

In particular, the outer regions of galaxies are expected to
hold a
wealth of information about the way in which galaxies are assembled. 
For example, in order to explain how the Milky Way was
formed, Eggen, Lynden-Bell \& Sandage (1962) and Searle \& Zinn (1978) studied the kinematics and
metallicity of stars and globular clusters in the bulge/halo of the Milky Way. 
While the former suggested the Galactic halo formed during a rapid
($\sim10^{8}$ years) monolithic collapse of a gaseous protocloud, the latter
 suggested formation via the accretion and merging of independent
 protogalactic fragments over a longer dynamical timescale (10$^{9}$ years). 
Importantly, Searle \& Zinn's (1978) seminal paper on Galactic globular clusters
constitutes the
basis for modern ideas relating to the formation of 
cosmological large scale structures in the universe (e.g. White \& Rees 1978).

Perhaps the main difference between the globular cluster systems of M31 and
the Milky Way is that the former is more populous with an estimated
total of $\sim$450 members (Barmby et al. 2000; Perrett et al. 2002; Huxor et
al. 2008; Caldwell et al. 2009) and may contain a significant population of
intermediate age (3--6 Gyr) globular clusters. Otherwise they share
many similarities: a Gaussian-like luminosity function that peaks around
M$_{\rm V}$ = --7.62 mag with a dispersion of 1.06 mag (Barmby, Huchra \& Brodie
2001); 
two subpopulations of mean metallicity [Fe/H] = --1.57 and
--0.61 associated with the halo and bulge, respectively (Barmby, Huchra \& Brodie 2001;
Forbes, Brodie \& Larsen 2001); and similar structural properties (Barmby et al.
2007). 

A relevant recent discovery is that of a metal-poor
stellar halo in M31 (Kalirai et al. 2006; Chapman et al. 2006).
At projected radii beyond 60 kpc Kalirai et al. (2006) estimate a mean halo
metallicity
of [Fe/H] = --1.26, which decreases to --1.48 if an $[\alpha/Fe]$
abundance ratio similar to the Milky Way's stellar halo is assumed. 
Beyond a projected radius of 70 kpc, outer halo globular clusters in M31 have
also been discovered recently but remain rare, these include 1 globular
cluster reported in Martin et al. (2006), 2 in Mackey et al.
(2007), 2 in Galleti et al. (2007) and 1 in Huxor et al. (2008).

In this Letter, we present Keck spectra for some of the furthest projected distant
($R_{\rm p}$) globular clusters known at this time in the halo of M31 --- GC5
and GC10 (Mackey et al. 2007) and GCM06 (Martin et al. 2006). 
Two of them (GC5 and GC10) do not
have spectroscopic information published to date. Thus, we present,
for the first time, their kinematics, structural parameters, ages and
spectroscopic metallicities. 
We briefly discuss our results within the context of the halo assembly of M31.

\section{Observations and Data Reduction}

The three globular clusters were selected from the imaging
survey reported by Martin et al. (2006, GCM06) and Mackey et al. (2007, GC5
and GC10).
The spectra were observed with the Keck/HIRES instrument (Vogt et al. 1994) on 2008
August 19. The instrument setup employed a
0\farcs86-wide, 7\farcs0-long slit, providing a spectral resolving power of
$R\approx50000$.
The cross-disperser angle was set to cover wavelengths
$\lambda\lambda=4020$--$8520$\,\AA. 
For each globular cluster, 3x900-s exposures were taken in succession, without
any intervening re-acquisition sequences, with the slit oriented at the
parallactic angle. Exposures were offset from each other in the
direction perpendicular to the parallactic angle (i.e.~along the horizon).
The first exposure was centred on the globular cluster with the others offset
either side of centre. The offsets were 1\farcs0 for GC5 and GCM06
and 0\farcs8 for GC10. The seeing was stable throughout the observations
at $\sim$0\farcs8. 

\begin{figure}
\includegraphics[width=88mm]{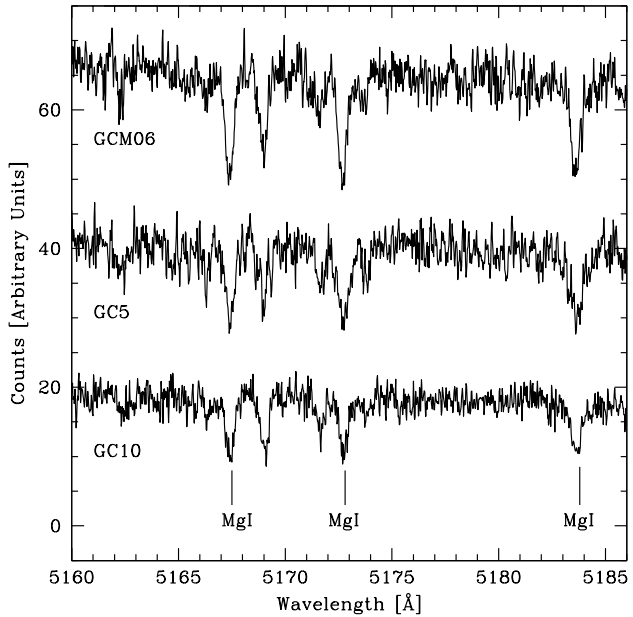}
 \caption{Integrated-light Keck/HIRES spectra of the 3 globular clusters around the Mg{\it b}
 Triplet wavelength region. The prominent absorption MgI lines are labeled just
 below the spectra. The typical signal-to-noise per pixel is 20 at this
 wavelength region.}
 \label{f:metal}
\end{figure}

Science and calibration data were reduced using the MAKEE (MAuna Kea
Echelle Extraction) Keck Observatory HIRES Data Reduction Program, 
written by Tom Barlow. Bias frames, flat-fields, order trace and ThAr arc images were taken
as part of the Keck base calibrations. The final reduction process comprised
bias-subtraction and flat-fielding, using base calibrations which were
previously co-added. 
The wavelength calibration was carried out with solutions
obtained from the ThAr arc exposures, which provided typical mean residuals
of 0.05 $\rm \AA$.  Spectra were then cosmic ray cleaned, sky-subtracted and
extracted into a series of heliocentric corrected 2D spectra with no
correction to a vacuum wavelength scale applied. The spectra were then combined
with inverse variance weighting to yield the
final spectrum of each cluster.
Basic properties for the 3 globular clusters are given in Table \ref{t:obs}.
Figure \ref{f:metal} displays a portion of the final spectra indicating the data
quality.

\begin{table}
\caption{Properties for the 3 outer globular clusters.}             
\label{t:obs}      
\centering                          
\begin{tabular}{lccc}      
\hline\hline                
Properties$^{*}$ & GCM06 & GC5 & GC10 \\   
\noalign{\vskip 0.1cm}
(1) & (2) & (3) & (4) \\ 
\hline                       

R.A. [J2000]	     &  00:50:42.5  &	       00:35:59.7   &  01:07:26.4   \\
Dec. [J2000]	     &  32:54:59.6  &	       35:41:03.6   &  35:46:49.7    \\
$E_{\rm B-V}$[mag]       &   0.08        & 0.08   & 0.09   \\
$r_{\rm h}$ [pc]     & 2.3 $\pm$ 0.20 & 	6.3 $\pm$ 0.15	    &  4.3 $\pm$ 0.15	       \\
$M_{\rm V,o}$ [mag]  & -8.5 $\pm$ 0.3  & 	-8.8	    &  -8.3	       \\ 
$R_{\rm p}$ [kpc]    & 100  	       &	78.5 	   &  99.9   \\

\hline   
\end{tabular}
\begin{minipage}{.88\hsize}
Notes.--- (*) As given in Martin et al. (2006) and Mackey et al. (2007).
\end{minipage}
\end{table}

\section{Data Analysis}

The galactocentric coordinates (X,Y) and projected distances
for our 3 outer halo globular clusters were calculated relative to an adopted
M31 central position of $\alpha_{\rm J2000}$ = 00$^{\rm h}$42$^{\rm
m}$44$^{\rm s}$.31, $\delta_{\rm J2000}$ = +41$^{\rm o}$16$^{\prime}$09\farcs4.
The position angle for the X-coordinate is 38$^{\rm o}$ (Kent 1989), and we
adopt 780 kpc as the distance to M31 (Holland 1998). 
At this distance, 1$^{\prime}$ corresponds to 228 pc.
The projected distances calculated in this way are similar to those quoted in
Table \ref{t:obs} from colour magnitude diagram (CMD) studies. These calculated radii
are used throughout this paper to facilitate comparison with other literature
work.

The velocity dispersions ($\sigma$) and heliocentric radial velocities
({\it v$_{\odot}$}) were obtained 
by fitting the observed spectra with the convolution of suitable template
spectra of giant stars taken with the same telescope and instrument and with a
Gaussian velocity profile using the program {\tt
pixfitgau} (see van
der Marel \& Franx 1993 for more details). The best fitting
parameters are obtained by $\chi$$^{2}$-minimisation procedures.
We used the spectral region 5150-5240 $\rm
\AA$, where strong features permit a reliable determination of the velocity
dispersion. The velocity dispersions were
also measured using the spectral region 5370-5420 $\rm \AA$, which contains many
iron lines. The results were consistent displaying differences of less than 1
km $s^{-1}$. The adopted values are based on the 5150-5240 $\rm \AA$ spectral
region.
The uncertainties quoted are only due to the random noise, that is, by taking
into account the CCD read-out noise per pixel ($\sim$ 0.6 e$^{-}$/pix) and
gain (1.9 e$^{-}$). 

The globular cluster masses were computed by using the Virial Theorem, being ${\it M}$ =
10$\times\sigma^{2}r_{\rm h}$/{\it G}, where $\sigma$ is the velocity
dispersion, $r_{\rm h}$ is the half-light
radius, {\it G} is the gravitational constant and a virial coefficient of 10 was
used (see Forbes et al. 2008). 
For GC5 and GC10 the $r_{\rm h}$ values were measured using images obtained with
the Advanced Camera for Surveys on Hubble Space Telescope (Mackey et al. 2007),
while for GCM06 it was obtained using ground-based
observations (Martin et al. 2006).

Spectroscopic metallicities were obtained by using metallicity
calibrations from Mg{\it b} and CH (Brodie
\& Huchra 1990; Perrett et al. 2002) and for Mg2 (Brodie \& Huchra 1990;
Buzzoni, Gariboldi \& Mantegazza 2002) indices. In addition, we also employed
the robust approach of Proctor, Forbes \& Beasley (2004), which performs the
simultaneous $\chi$$^{2}$-minimisation of a large number of 
spectral indices.  This method 
leverages the entire suite of available indices to estimate stellar 
population parameters and, in particular, makes apparent those indices 
which deviate from the general trend of the data, effectively 
eliminating errors in calibration or the spectra themselves (e.g. 
sky-line contamination, bad pixels etc). 

The Single Stellar Population (SSP) models used here are from Thomas, Maraston
\& Korn (2004), which have been used to recover reliable ages 
and metallicities for both Galactic and extragalactic globular clusters (e.g.
Mendel, Proctor \& Forbes 2007; Beasley et al. 2008).
However, we note that such an analysis is usually applied to much lower
resolution data than obtained here. 
Hence, absorption line indices were measured after broadening the spectra to the 
wavelength dependent Lick/IDS system resolution as described in 
Worthey \& Ottaviani (1997). 
The Lick index definitions were taken from Worthey \& Ottaviani (1997) and 
Trager et al. (1998). 

\section{Results and Discussion}

We derive a heliocentric radial velocity of $-$358.3 $\pm$ 1.9 km s$^{-1}$ for
GC5, {\it v}$_{\rm \odot}$ = $-$291.2 $\pm$ 2.1 km s$^{-1}$ for GC10 and a value of
$-$354.7 $\pm$ 2.2 km s$^{-1}$ for GCM06. All 3 globular clusters are consistent
with being part of the M31 ({\it v}$_{\rm \odot}$ = --300 km s$^{-1}$) globular cluster
system.
Recently, using a low-resolution spectrum (R $\sim$ 1300), Galleti et
al. (2007) determined {\it v}$_{\rm \odot}$ = $-$312 $\pm$ 17 km s$^{-1}$ for GCM06.
This mean difference between the heliocentric
radial velocities suggests the errors quoted in the lower resolution work may
have been underestimated.

\begin{figure}
\includegraphics[width=88mm]{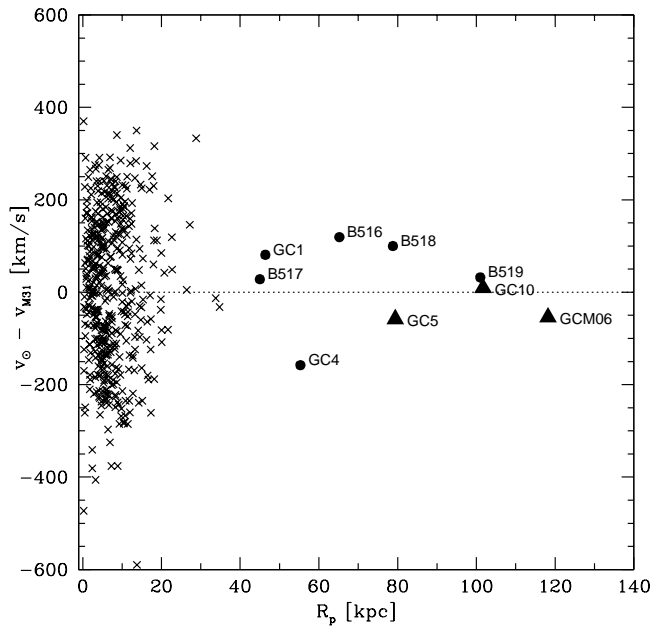}
\caption{Heliocentric radial velocities for M31 globular clusters corrected to
the M31 velocity frame against the projected radius.
Symbols are as follows: the {\it crosses} for globular clusters at $R_{\rm p}$
$\leq$ 40 kpc, while the {\it solid symbols} are for those at large projected
distances ($R_{\rm p}$ $\geq$ 40 kpc). The present
sample is represented by {\it solid
triangles}, while the outer globular clusters of Galleti et al. (2007)
and Mackey et al. (2007) are represented by the {\it solid circles}.}
\label{f:vel}
\end{figure}

As shown in Fig. \ref{f:vel} our radial velocity results are clustered around
M31's systemic recession velocity of $-$300 km s$^{-1}$. 
Regarding this issue, Chapman et al. (2006) analysed a large sample of red
giant branch stars in the halo of M31 (between 10 $\leq$ $R_{\rm p}$ $\leq$
70 kpc) and fitting a simple model to the observed data they obtained the
dispersion of the radial velocity as a function of the
projected radius $R_{\rm p}$\footnote{$\sigma_{v_{\odot}}$($R_{\rm p}$) = 152 -
0.90$R_{\rm p}$ km s$^{-1}$, with $R_{\rm p}$ given in kpc.}. Their linear relation
implies $\sigma_{v_{\odot}}$ = 62 km s$^{-1}$ at 100 kpc
distance from the centre of M31. 
Combining the velocity results for the 5 globular clusters beyond
$R_{\rm p}$ = 70 kpc presented in Fig. \ref{f:vel}, we derive a velocity
dispersion of 65 km s$^{-1}$ in the outer halo of M31.
For the Milky Way globular cluster system, Battaglia et al. (2005) found that
the radial velocity dispersion shows a nearly constant value of 120 km s$^{-1}$ out to 30 
kpc, which declines down to 50 km s$^{-1}$ at about 120 kpc. 
The declining velocity dispersion in the outer halo suggests a relatively low
mass halo, in which the virial velocity of the halo is less than the
rotation speed of the disk (Abadi, Navarro \& Steinmetz 2006), similar to that
proposed for the Milky Way by Klypin, Zhao \& Somerville (2002).

\begin{figure}
\includegraphics[width=88mm]{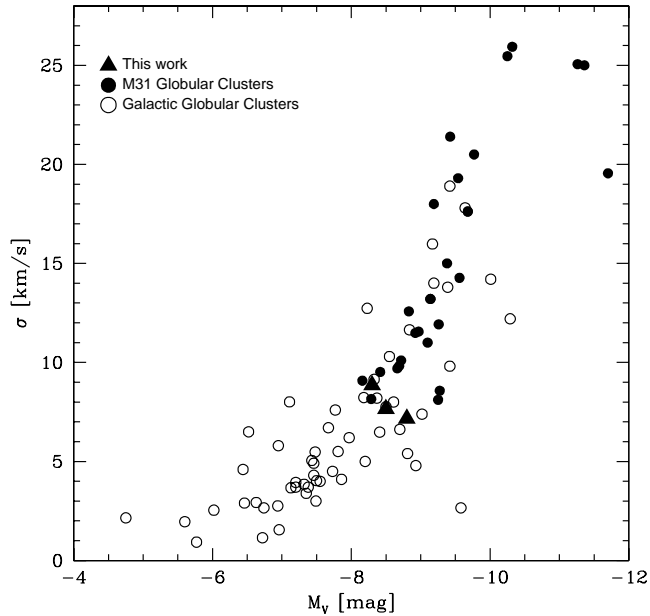}
 \caption{Internal velocity dispersion as a function of the absolute visual magnitude for
 the present sample ({\it triangles}), Galactic globular clusters ({\it
 open circles}) and for other globular clusters in
 M31 ({\it solid circles}). Refer to the text for references.}
  \label{f:veldisp}
\end{figure}

In Fig. \ref{f:veldisp} we show the measured globular cluster internal velocity
dispersion versus the absolute visual magnitude.
It is evident that our 3 outer globular clusters resemble those found for
Galactic globular clusters (Pryor \& Maylan 1993) and also for other globular
clusters in M31 (Barmby et al. 2007). 
Such a scaling relation is not well understood to date. Yet, as briefly discussed by
Djorgovski (1991, 1993), the primordial $L$-$\sigma$ relation (slope $\sim$1) of
young globular clusters would be affected by subsequent dynamical
processes (tidal shocks, for example), which would lead to the $L$-$\sigma$
power-law relation observed by the present day.

For our sample, the dynamical mass-to-light ratio in the V band $M/L_{\rm V}$ ranges from
1.50 to 4.51 (in solar units). Our results are in agreement with those obtained
by Dubath \& Grillmair (1997) for nine M31 globular clusters (1.0 $\leq$ $M/L_{\rm V}$
$\leq$ 6.5). 
In addition, Pryor \& Meylan (1993)
obtained a mean dynamical $M/L_{\rm V}$ = 2.3 $\pm$ 1.1 for the Galactic
globular cluster system. 
Hence, within the uncertainties, GCM06 and GC5 display typical
Galactic globular cluster $M/L_{\rm V}$ ratios, while GC10 is rather on the
high side. 

\begin{table}
\setlength\tabcolsep{2.5pt}
\centering
\begin{minipage}{140mm}
\caption{Derived parameters for the 3 outer globular clusters.}
\label{t:dynamics}

\begin{tabular}{lccccc}
\hline\hline
\noalign{\vskip 0.1cm}  
\rm 
ID & R$_{\rm p}$ & $\sigma$ &  M & M/L$_{\rm V}$ & Age\\
\noalign{\vskip 0.1cm}  
  &  [kpc] & [$\rm km s^{-1}$] &  [10$^{5}$$M_{\rm \odot}$] &  [$M_{\rm \odot}/L_{\rm \odot}$] & [Gyr] \\
\noalign{\vskip 0.1cm}  
(1)    & (2) & (3) & (4) & (5)& (6)\\

\noalign{\smallskip}
\hline
\noalign{\vskip 0.1cm}
\hbox{GCM06}	  & 118.2  & 7.7 $\pm$ 0.4 & 3.14 $\pm$ 1.57 & 1.50 $\pm$ 0.75 & 7.1  $\pm$ 3.0  \\
\hbox{GC5}  	  & 79.4   & 7.2 $\pm$ 0.4 & 7.51 $\pm$ 1.50 & 2.73 $\pm$ 0.54 & 10.0 $\pm$ 3.0\\
\hbox{GC10} 	  & 101.7  & 8.8 $\pm$ 0.7 & 7.83 $\pm$ 1.56 & 4.51 $\pm$ 0.90 & 12.6 $\pm$ 3.0\\

\hline	
\noalign{\smallskip}										      
\end{tabular}
									      
\end{minipage}
\end{table}

When compared to the Galactic globular cluster system, one remarkable aspect of
the M31 globular cluster system is its large range in age. That is, while the
Milky Way presents a very homogeneous old population, M31
displays a young population ($\leq$ 2 Gyr), a number of intermediate
age globular clusters (3-6 Gyr) and old globular clusters ($\geq$ 7 Gyr) as
well (e.g. Beasley et al. 2005). Based on our stellar population analysis,
we found that the present outer halo globular clusters of M31 have a mean
age of around 10 $\pm$ 3 Gyr.  
This mean spectroscopic age determination is in agreement with the
color-magnitude diagrams presented in Martin et al. (2006) and Mackey et
al. (2007) for these globular clusters. Furthermore, our dynamical
mass-to-light ratios ($M/L_{\rm V}$ $\geq$ 1.5) are compatible with an old
stellar population.
Our final derived parameters are summarized in Table
\ref{t:dynamics}.

Spectroscopic metallicities for outer metal-poor globular clusters are
essential to the scenarios of chemical enrichment of the halo in galaxies since
these objects are considered halo-tracers. In Table \ref{t:metal} we present 
our spectroscopic metallicity determinations and, for comparison,
the photometric metallicities.
The final metallicities were computed as the
median of the different spectroscopic values presented in Table \ref{t:metal}, with a robust
uncertainty of 1.5$\times$SIQR (Semi-InterQuartile Range) being adopted.
This lead to [Fe/H] = $-$1.37 $\pm$ 0.15 dex for GCM06, [Fe/H] = $-$1.33 $\pm$ 0.12 dex
for GC5 and [Fe/H] = $-$1.73 $\pm$ 0.20 dex for GC10. These values, taking all uncertainties into
account, are consistent with the CMD ones reported in Table \ref{t:metal} for GC10 and
GCM06, while GC5 is a factor of 3 more metal-rich than the CMD
estimate. 
Note, however, that although the metallicity quoted for
GC5 based upon the {\it Mgb} method is also supported by visual inspection of
the high-quality spectra shown in Fig. \ref{f:metal}, the higher value
obtained by the median of the different spectroscopic measurements was adopted.

Both the photometric and spectroscopic methods have their advantages and disadvantages.
Whilst the former is reddening and distance-modulus dependent, the latter is calibration and model-dependent. 
Furthermore, photometric metallicities are also susceptible to the set of
isochrones used as well as to age-metallicity degeneracy on the red
giant branch. 
On the other hand, measurements of age, metallicity and elemental abundance
from integrated spectra using a single index or pairs of 
orthogonal indices (e.g. H$\beta$ vs. [MgFe]) are especially 
susceptible to index calibration errors, a particular concern for our 
uncalibrated data.

\begin{table}
\centering
\caption{Metallicity determinations for the 3 outer globular clusters.}
\label{t:metal}
\begin{tabular}{lccccc}
\hline\hline

\hbox{Method} &  \hbox{GCM06} & \hbox{GC5} & \hbox{GC10} \\
\noalign{\smallskip}
\hline
\noalign{\smallskip}
\rm 

$\chi^{2}$     &  $-$1.65 $\pm$ 0.10 & $-$1.40 $\pm$  0.10 & $-$1.73 $\pm$ 0.10   \\
Mgb$^{*,a}$    &  $-$1.56 $\pm$ 0.11 & $-$1.74 $\pm$ 0.08 & $-$1.82 $\pm$ 0.08   \\
CH$^{*,b}$     &  $-$1.37 $\pm$ 0.25 & $-$1.24 $\pm$ 0.25 & $-$1.81 $\pm$ 0.24   \\ 
Mg2$^{*,c}$    &  $-$1.36 $\pm$ 0.05 & $-$1.33 $\pm$ 0.06 & $-$1.54 $\pm$ 0.01   \\
CMD$^{d}$      &  $-$1.30 $\pm$ 0.15 & $-$1.84 $\pm$ 0.15 & $-$2.14 $\pm$ 0.15   \\
               &   &  &   \\
Adopted        &  $-$1.37 $\pm$ 0.15 & $-$1.33 $\pm$ 0.12 & $-$1.73 $\pm$ 0.20   \\

\hline	
\noalign{\smallskip}										      
\end{tabular}	
\begin{minipage}{.88\hsize}
 Notes.--- (*): Average values using different calibrations as follows : (a),(b): Brodie \& Huchra
(1990) and Perrett et al. (2002); (c): Brodie \& Huchra (1990) and Buzzoni et al.
(1992); (d): Isochrone fitting from Martin et al. (2006) and Mackey et al. (2007).\\
\end{minipage}									      
\end{table}

\begin{figure}
\includegraphics[width=88mm]{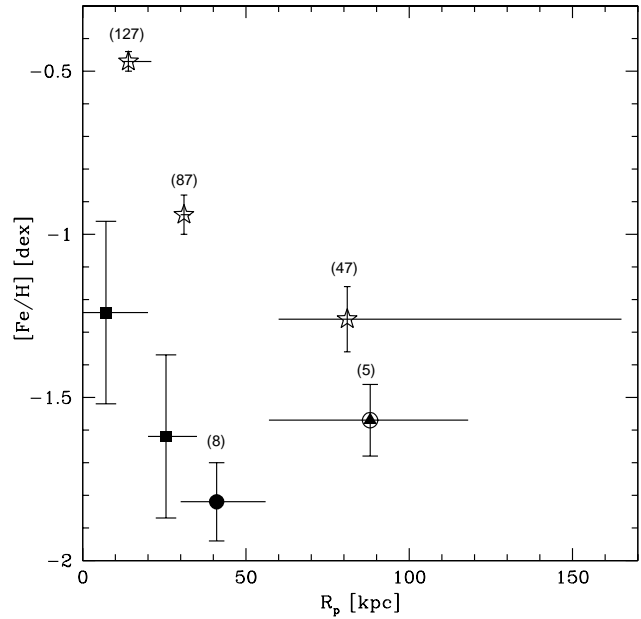}
 \caption{Mean metallicity plotted against the projected
 radius. Four different symbols are used: 
 (i) red giant branch stars ({\it open stars}) from Kalirai et al. (2006);
 (ii) metal-poor globular clusters ({\it solid squares}) as analysed in Lee et
 al. (2008); (iii) metal-poor globular clusters ({\it solid circle}) from Mackey
 et al. (2006; 2007) and Galetti et al. (2007); (iv) metal-poor globular clusters ({\it solid circled
 triangle}) from this work and combined with those presented in Mackey et al.
 (2006) and Galleti et al. (2007) for EC4 ([Fe/H] = --1.84) and B518
 ([Fe/H]= --1.6), respectively. Error bars correspond to the standard error on the mean [Fe/H] measurements.
 When available, the total number of objects used is also indicated. 
 The horizontal lines delineate the bin-size adopted.}
\label{f:grad}
\end{figure}

In Fig. \ref{f:grad} we present the behaviour of the mean metallicity as a
function of the projected radii in the halo of M31 using our metal-poor
globular cluster sample, together with additional values from the
literature for field stars and globular clusters.
Three main aspects are illustrated in this figure. 
Firstly, there seems to exist a
metallicity gradient for the metal-poor globular clusters at a projected distance less
than 50 kpc from the centre of M31. 
Secondly, we find tentative evidence for the mean
globular cluster metallicity to flatten off at [Fe/H] $\sim$ --1.6, which is similar to the
lower threshold seen in many other galaxies (e.g. Forbes et al. 2000).
Thirdly, our results confirm that the globular clusters in
M31 are systematically more metal-poor than their outermost counterpart
field stars, bearing in mind that we may be sampling different physical radii.
This point can be explained in the context of chemical
pre-enrichment having taken place in M31 before most of the halo formation has
occurred (Durrell, Harris \& Pritchet 1994, 2001). 
Similar results have also been found for
giant elliptical galaxies (e.g. Harris, Harris \& Poole 1999). 
However, Schorck et al. (2008) show that the metallicity
distribution function of the Galactic halo field stars is 
indistinguishable from that of the Galactic globular cluster system.

Currently, it is not clear what fraction of the
Milky Way's halo has been assembled via dissipational collapse
(monolithic collapse and gas rich mergers), and which by dissipationless
accretion. However, the global properties of the present-day
globular clusters of the Milky Way suggest a scenario where the inner halo
(within 10 kpc) was formed via dissipative collapse, while the outer
halo has an accretion origin (e.g. Bica et al. 2006). 
The metallicity trend in Fig. \ref{f:grad} is also consistent with this
picture. It suggests that the steep metallicity
gradient for the globular cluster system of M31 in the
inner halo flattens in the outer
regions. Such a flattening could be indicative of the
outer halo globular clusters being likely accreted from satellites during an
early period of merging (see e.g. Font et al. 2006).

\section{Summary}

In summary, we confirm that the 3 outer globular clusters in the present sample
are kinematically members of the M31 globular cluster system and have structural
parameters that, within the uncertainties, resemble those of the Milky Way globular clusters.
Furthermore, they are characterized by an old
stellar population ($>$ 7 Gyr), whose mean
metallicity matches the M31 metallicity peak for halo globular
clusters (e.g. Barmby, Huchra \& Brodie 2001). 
Unlike the Milky Way, we confirm that the outer halo globular clusters
of M31 are systematically more metal-poor than their counterpart field stars
(Kalirai et al. 2006) at a given projected radius.
We also find evidence for a declining velocity dispersion with radius for
the globular cluster system.

We note that the most distant globular cluster in the Milky Way, AM~1, lies
at 120 kpc.
It is reasonable to expect that more globular clusters in the outer
halo of M31 will be discovered with time, which should be followed-up with
integrated-light
spectra.

\section*{Acknowledgments}

We would like to thank Adrian Malec for his support with the MAuna Kea Echelle
Extraction Program on MAC Systems and S$\ddot{\rm o}$eren Larsen for sending us
template star observations and useful comments. We are grateful
to Jay Strader and the anonymous referee for constructive comments and
suggestions.
AAB acknowledges CNPq for financial support 200227/2008-4 (PDE).  
DF and GKTH thank ARC for financial support.

\end{document}